\documentclass{article}
\usepackage{amssymb}

\input{tcilatex}
\begin{document}

\date{}
\author{\textbf{Howard E. Brandt} \\
}
\title{Maximal Proper Acceleration and the Quantum-to-Classical Transition }
\maketitle

\begin{abstract}
I first review the physical basis for the universal maximal proper
acceleration. Next, I introduce a new formulation for a relativistic scalar
quantum field which generalizes the canonical theory to include the limiting
proper acceleration. This field is then used to construct a simple model of
an uncorrelated many-body system. I next argue that for a macroscopic object
consisting of more than Avogadro's number of atoms, any supposed quantum
state of the object is negligibly small, so that for all practical purposes,
the object is best described by classical mechanics. Thus, a new explanation
is offered for the quantum-to-classical transition and the absence of
quantum superposition of macroscopic objects in the everyday world.

\end{abstract}

\section{INTRODUCTION}

There is no generally accepted theory of why the every day world of
macroscopic objects is not usefully described in terms of quantum states.
For example, a planet is never observed to be in a quantum superposition
state. It has however been speculated that for the description of a
classical macroscopic many-body system of sufficient complexity, quantified
by the number of atoms of which it is composed, quantum mechanics can be
replaced by classical mechanics. [Of course for a highly correlated
mesoscopic system such as a Bose condensate, which consists of a single
quantum state, a quantum description is needed. Also, quantum mechanics is
clearly needed to understand the atomic and molecular structure of
macroscopic objects.] It is here to be argued that for ordinary macroscopic
objects consisting of more than Avogadro's number of atoms, any supposed
quantum state of the object as a whole is negligibly small, so that for all
practical purposes the object is best described by classical mechanics. This
follows from a natural extension of Lorentz-invariant quantum field theory
to include the physics-based upper bound on physically possible proper
accelerations \cite{B1}-\cite{B5}.

\section{MAXIMAL PROPER ACCELERATION}

Heuristic arguments are first given for the existence of a universal upper
bound on proper acceleration. In the presence of purely gravitational
fields, macroscopic particles follow geodesic paths with vanishing proper
acceleration. In the presence of non-gravitational forces, the proper
acceleration is nonvanishing, however it follows from elementary physical
reasoning that there is a maximum possible proper acceleration, the
so-called maximal proper acceleration relative to the vacuum \cite{B1}. The
physical basis for maximal proper acceleration is direct \cite{B5}. By the
time-energy uncertainty principle, virtual particle-antiparticle pairs of
mass $m$ occur in the vacuum during a time $\hbar /2mc^{2}$ and over a
distance $\hbar /2mc$, the Compton wavelength of the particles. This is the
ordinary vacuum polarization. Here $\hbar $ is Planck's constant divided by $%
2\pi $, and $c$ is the velocity of light in vacuum. In an accelerated frame,
the inertial force acts on such a virtual particle in the vacuum
polarization, and if an amount of energy equal to its rest energy is
imparted to it, the particle becomes real. The inertial force with magnitude 
$ma$ on a virtual particle having proper acceleration $a$ acts over a
distance within which the particle can be created, namely, of the order of
its Compton wavelength $\hbar /2mc$, thereby doing work of order $(ma)(\hbar
/2mc)$. This follows from the fact that the proper acceleration is the
magnitude of the ordinary acceleration in the instantaneous rest frame of
the particle. If this work is equated to the rest energy $mc^{2}$ of the
particle, it then follows that, for proper acceleration of order%
\begin{equation}
a_{M}\sim 2mc^{3}/\hbar ,
\end{equation}%
particles of mass $m$ are copiously produced out of the vacuum. The larger
the acceleration, the larger are the masses of the created particles. In the
extreme, if the acceleration is sufficiently large, the created particles
will be black holes. For this to occur, the size of \ a created particle
(namely, of the order of its Compton wavelength $\hbar /2mc$) must be less
than its Schwarzschild radius $2Gm/c^{2}$, where $G$ is the universal
gravitational constant. Thus the minimum possible mass of a black hole\ is
of the order of the Planck mass $(\hbar c/G)^{1/2}$ \cite{B1}. Next, if the
Planck mass $(\hbar c/G)^{1/2}$ is substituted in Eq. (1), it follows that,
for proper acceleration $a_{M\text{ }}$given by%
\begin{equation}
a_{M}=2\pi \alpha (c^{7}/\hbar G)^{1/2},
\end{equation}%
where $\alpha $ is a number of order unity, there will be copious production
of Planck mass black holes out of the vacuum, resulting in the formation of
a manifest spacetime foam and the topological breakdown of the classical
spacetime structure, as well as the breakdown of the very concept of
acceleration \cite{B1}, \cite{B2},  \cite{B5}. Thus Eq. (2) gives the
maximum possible proper acceleration relative to the vacuum.

\section{RELATIVISTIC QUANTUM FIELDS}

There follows a new natural extension of Lorentz-invariant quantum field
theory to include the physics-based upper bound on physically possible
proper accelerations. In Minkowski spacetime with a negative-signature, the
spacetime line element (interval), is given by%
\begin{equation}
ds^{2}=dx_{\mu }dx^{\mu
}=dx_{0}^{2}-dx^{2}-dy^{2}-dz^{2}=c^{2}dt^{2}-dx^{2}-dy^{2}-dz^{2}\geq 0,
\end{equation}%
which follows directly from the fact that the velocity of any object cannot
exceed the velocity $c$ of light. Here $x^{0}=x_{0}=ct$ where $t$ is the
time, and $\{x^{\mu }\}=\{x^{0},x^{1},x^{2},x^{3}\}=\{ct,x,y,z\}=\{x^{0},%
\mathbf{x}\}$ are the coordinates of a point in spacetime. In the present
work, the $x^{\mu }$ are also taken to be the coordinates of a macroscopic
measuring device making measurements of a field or detecting a particle at
the point $x^{\mu }$ in spacetime. Associated with the Minkowski line
element is the four-dimensional d'Alembertian operator

\bigskip 
\begin{equation}
\square ^{(4)}\equiv \frac{\partial ^{2}}{\partial x_{\mu }\partial x^{\mu }}%
,
\end{equation}%
which appears in the Klein-Gordon equation for a relativistic scalar quantum
field $\phi (x)$ describing particles of mass $m$, namely 
\begin{equation}
\left( -\hbar ^{2}\frac{\partial ^{2}}{\partial x_{\mu }\partial x^{\mu }}%
-m^{2}c^{2}\right) \phi =0,
\end{equation}%
or equivalently%
\begin{equation}
\left( \hbar ^{2}\square ^{(4)}+m^{2}c^{2}\right) \phi =0.
\end{equation}%
The four-velocity of a macroscopic measuring device, moving relative to a
particle excitation of the quantum field at point $x^{\mu }$ and measuring
the particle, is given by

\begin{equation}
v^{\mu }=dx^{\mu }/ds.
\end{equation}%
It is to be emphasized here that $v^{\mu }$ is not the four-velocity of the
microscopic quantum particle \ [Since the measured particle is localized at
a point in spacetime, its four-velocity is indeterminate due to the quantum
uncertainty principle.] Also, the four-acceleration $A^{\mu }$ of the
measuring device is%
\begin{equation}
A^{\mu }=dv^{\mu }/ds.
\end{equation}%
The corresponding proper acceleration $A$ of the measuring device is defined
by%
\begin{equation}
A^{2}=-c^{4}\frac{dv_{\mu }}{ds}\frac{dv^{\mu }}{ds},
\end{equation}%
which follows from the fact that proper acceleration is the magnitude of the
ordinary acceleration in the instantaneous rest frame of an object. Under
the constraint of the universal upper bound $a_{M}$ on proper acceleration,
one requires that%
\begin{equation}
A^{2}\leq a_{M}^{2}.
\end{equation}%
It is useful to define%
\begin{equation}
\rho _{0}\equiv \frac{c^{2}}{a_{M}},
\end{equation}%
which according to Eq.\ (2) is of the order of the Planck length. From Eqs.
(9)-(11), it follows that

\begin{equation}
-c^{4}\frac{dv_{\mu }}{ds}\frac{dv^{\mu }}{ds}\leq \rho _{0}^{-2}.
\end{equation}%
Equivalently then%
\begin{equation}
d\sigma ^{2}\equiv ds^{2}+\rho _{0}^{2}dv_{\mu }dv^{\mu }\geq 0,
\end{equation}%
or%
\begin{equation}
d\sigma ^{2}\equiv c^{2}dt^{2}-dx^{2}-dy^{2}-dz^{2}+\rho
_{0}^{2}(dv_{0}^{2}-dv_{x}^{2}-dv_{y}^{2}-dv_{z}^{2})\geq 0,
\end{equation}%
where $v_{0},v_{x},v_{y},$and $v_{z}$ are the time and spatial components of
the four-velocity. Equation (14), expressing the fact that there is a
maximum possible proper acceleration, is here taken to be the line element
in the eight-dimensional spacetime-four-velocity space \ (tangent bundle of
Minkowski spacetime \cite{B3}, \cite{B4}). This is directly analogous to the
fact that the ordinary spacetime line element Eq.(3) follows from the
maximum possible velocity. Next, just as the four-dimensional d'Alembertian
operator $\square ^{(4)}$ is associated with the line element Eq. (3), one
can take the eight-dimensional operator $\square ^{(8)}$ defined by 
\begin{equation}
\square ^{(8)}\equiv \frac{\partial ^{2}}{\partial x_{\mu }\partial x^{\mu }}%
+\frac{1}{\rho _{0}^{2}}\frac{\partial ^{2}}{\partial v_{\mu }\partial
v^{\mu }},
\end{equation}%
as the operator associated with the line element Eq. (14). [Equation (15) is
the Laplace-Beltrami operator on the tangent bundle of Minkowski spacetime 
\cite{B6}] Next, by analogy with Eq. (5), one naturally defines a
generalized scalar quantum field $\phi (x,v)$ satisfying%
\begin{equation}
\square ^{(8)}\phi (x,v)\equiv 0,
\end{equation}%
or equivalently,%
\begin{equation}
\left( \frac{\partial ^{2}}{\partial x_{\mu }\partial x^{\mu }}+\frac{1}{%
\rho _{0}^{2}}\frac{\partial ^{2}}{\partial v_{\mu }\partial v^{\mu }}%
\right) \phi (x,v)=0.
\end{equation}%
Equation (17) can also be written as%
\begin{equation}
\left( \square _{x}+\rho _{0}^{-2}\square _{v}\right) \phi (x,v)=0,
\end{equation}%
where the spacetime and four-velocity d'Alembertian operators are defined by%
\begin{equation}
\square _{x}\equiv \frac{\partial ^{2}}{\partial x_{\mu }\partial x^{\mu }},
\end{equation}%
and%
\begin{equation}
\square _{v}\equiv \frac{\partial ^{2}}{\partial v_{\mu }\partial v^{\mu }},
\end{equation}%
respectively.

Next consider a possible separable single-mode solution $\phi (x,v)\equiv
\phi (x^{\mu },v^{\mu })$ to Eq. (18) of the form 
\begin{equation}
\phi (x,v)=\phi _{1}(x)\phi _{2}(v),
\end{equation}%
in which the dependence on the spacetime coordinates $x^{\mu }$ is separated
from the dependence on the four-velocity coordinates $v^{\mu }$. If one
substitutes Eq. (21) in Eq. (18), then for non-vanishing $\phi _{1}(x)$ and $%
\phi _{2}(v)$, one obtains 
\begin{equation}
\frac{\square _{x}\phi _{1}(x)}{\phi _{1}(x)}+\rho _{0}^{-2}\frac{\square
_{v}\phi _{2}(v)}{\phi _{2}(v)}=0.
\end{equation}%
Since the first term of Eq. (22) depends only on $x$, and the second term
depends only on $v$, both terms must be given by constants with the same
absolute value, but with opposite signs. The constants can be defined in
complete generality by $\pm (\mu c/\hbar )$, in which the constant $\mu $ is
at this point an arbitrary constant to be determined. One therefore has%
\begin{equation}
\frac{\square _{x}\phi _{1}(x)}{\phi _{1}(x)}=-\left( \frac{\mu c}{\hbar }%
\right) ^{2}
\end{equation}%
and%
\begin{equation}
\rho _{0}^{-2}\frac{\square _{v}\phi _{2}(v)}{\phi _{2}(v)}=\left( \frac{\mu
c}{\hbar }\right) ^{2}.
\end{equation}%
Possible solutions to Eqs. (23) and (24) are given by%
\begin{equation}
\phi _{1}^{\pm }(x)=\phi _{10}e^{\pm ik\cdot x},
\end{equation}%
\begin{equation}
\phi _{2}^{\pm }(v)=\phi _{20}e^{\mp q\cdot v}\theta \left( \pm q\cdot
v\right) ,
\end{equation}%
respectively, where $\phi _{10}$ and $\phi _{20}$ are constants, $k\cdot
x\equiv k_{\mu }x^{\mu }$, $q\cdot v\equiv q_{\mu }x^{\mu }$, and $k^{\mu }$
and $q^{\mu }$ are Lorentz four-vectors, still to be determined. Also in
Eq.\ (26), $\theta \left( x\right) $ is the Heaviside step function defined
by%
\begin{equation}
\theta \left( x\right) =\left\{ 
\begin{array}{c}
\ 1,\ x\geq 0 \\ 
0,\ \ x,0%
\end{array}%
\ \right. .
\end{equation}%
For $\phi _{2}^{\pm }(v)$ in Eq. (26) to be bounded for $|q\cdot
v|\rightarrow \infty $, and the exponent to be decreasing, the negative sign
must be chosen in the exponent fort $q\cdot v>0,$ and the positive sign must
be chosen for $q\cdot v<$0. This is insured by the appearance of the
Heaviside function $\theta \left( \pm q\cdot v\right) $ in Eq. (26)$.$ Next
substituting Eqs. (25) and (26) in Eqs. (23) and (24), respectively, one
obtains \ 
\begin{equation}
k^{2}=\left( \frac{mc}{\hbar }\right) ^{2},
\end{equation}%
and%
\begin{equation}
\rho _{0}^{-2}q^{2}=\left( \frac{mc}{\hbar }\right) ^{2},
\end{equation}%
respectively. Next, Eq. (23) can be rewritten as follows:%
\begin{equation}
\hbar ^{2}\square _{x}\phi _{1}(x)+\mu ^{2}c^{2}\phi _{1}(x)=0,
\end{equation}%
which is recognized to be the standard Klein-Gordon equation for a scalar
quantum field describing particles of mass%
\begin{equation}
\mu =m.
\end{equation}%
Thus, substituting Eq. (31) in Eq.(30), one has%
\begin{equation}
\hbar ^{2}\square _{x}\phi _{1}(x)+m^{2}c^{2}\phi _{1}(x)=0.
\end{equation}%
Lorentz invariance requires that the four-momentum of the particle must
satisfy the standard relativistic relation between energy, momentum, and
mass, namely,%
\begin{equation}
p^{2}=p_{\mu }p^{\mu }=p_{0}^{2}-|\mathbf{p}|^{2}=m^{2}c^{2},
\end{equation}%
where $p^{\mu }$ is the four-momentum of the particle, with spatial component%
$\ \mathbf{p}$ and time component $p^{0}$. Also, substituting Eq. (25) in
Eq, (32), then%
\begin{equation}
k^{\mu }=\frac{p^{\mu }}{\hbar }.
\end{equation}%
In earlier work \cite{B6}, \cite{B8}, \cite{B9}, the four-vector $q^{\mu }$
in Eq. (26) was taken to be given by $q^{\mu }=\rho _{0}p^{\mu }/\hbar ,$
which clearly satisfies Eq. (29). It was also pointed out that there are
other possible choices for $q^{\mu }$ which one might consider \cite{B9}. In
the present work, the following new choice is made for the four-vector $%
q^{\mu }$:%
\begin{equation}
q^{\mu }=\frac{\rho _{0}}{\lambda _{0}}\frac{a^{\mu }}{a},
\end{equation}%
where 
\begin{equation}
\lambda _{0}=\frac{\hbar }{mc}
\end{equation}%
is the wavelength for a particle at rest with rest mass $m$, $a^{\mu }$ is
the four-acceleration of the particle, and $a$ is the corresponding proper
acceleration of the particle, namely,%
\begin{equation}
a^{2}=a_{0}^{2}-|\mathbf{a}|^{2},
\end{equation}%
where $\mathbf{a}$ is the spatial component of four-acceleration, and $%
a^{0}=a_{0}$ is the time component. [Note that in standard quantum field
theory, the momentum, velocity, and acceleration of a particle at a specific
spatial location lacks meaning because of the uncertainty principle;
however, in quantum field theory the particle momentum is merely a
parameter, and in the present work the particle acceleration is also a
parameter. Also, it is well to recall that in the Bohm interpretation of
quantum mechanics, the particle position, velocity, and acceleration of a
particle at a point are simultaneously meaningful \cite{L1}, \cite{DT1}, 
\cite{H1}.] Next one notes that the four-vector%
\begin{equation}
n^{\mu }\equiv \frac{a^{\mu }}{a}
\end{equation}%
is a unit vector, namely, 
\begin{equation}
n^{2}=n_{\mu }n^{\mu }=\frac{a^{2}}{a^{2}}=1,
\end{equation}%
and using Eqs. (35), (38), and (39), one has%
\begin{equation}
q^{2}=q_{\mu }q^{\mu }=\frac{\rho _{0}^{2}}{\lambda _{0}^{2}}=\left( \frac{%
\rho _{0}mc}{\hbar }\right) ^{2},
\end{equation}%
which agrees with Eq. (29). \ Next substituting Eqs. (34) and (35 ),
respectively, in Eqs. (25) and (26), respectively, one obtains%
\begin{equation}
\phi _{1}^{\pm }(x)=\phi _{10}e^{\pm ip\cdot x/\hbar },
\end{equation}%
and%
\begin{equation}
\phi _{2}^{\pm }(v)=\phi _{20}e^{\mp \frac{\rho _{0}}{\lambda _{0}}\frac{%
a\cdot v}{a}}\theta \left( \pm \frac{a\cdot v}{a}\right) ,
\end{equation}%
respectively, where 
\begin{equation}
p\cdot x=p_{\mu }x^{\mu },
\end{equation}%
and%
\begin{equation}
a\cdot v=a_{\mu }v^{\mu }.
\end{equation}

Next, substituting Eq. (42) in Eq. (24), one obtains%
\begin{equation}
\square _{v}\phi _{2}^{\pm }(v)=\phi _{20}\left( \frac{\rho _{0}}{\lambda
_{0}}\right) ^{2}\left[ \delta ^{\prime }\left( \pm \frac{a\cdot v}{a}%
\right) -2\delta \left( \pm \frac{a\cdot v}{a}\right) +\theta \left( \pm 
\frac{a\cdot v}{a}\right) \right] e^{\mp \frac{\rho _{0}}{\lambda _{0}}\frac{%
a\cdot v}{a}},
\end{equation}%
where $\delta (x)$ is the Dirac delta function. From Eq. (37), one has%
\begin{equation}
\{a^{\mu }\}=\{a_{0},\mathbf{a}\}=\left\{ \left( a^{2}+\left\vert \mathbf{a}%
\right\vert ^{2}\right) ^{1/2},\mathbf{a}\right\} .
\end{equation}%
Also, using Eqs. (7) and (3), one has%
\begin{equation}
\{v^{\mu }\}=\left\{ \frac{dx^{0}}{ds},\frac{d\mathbf{x}}{ds}\right\}
=\left\{ \gamma ,\gamma \frac{d\mathbf{x}/dt}{c}\right\} ,
\end{equation}%
where%
\begin{equation}
\gamma =\left( 1-(\frac{d\mathbf{x}/dt}{c})^{2}\right) ^{-1/2}.
\end{equation}%
Using Eqs.\ (46) and (47), it can be shown that [See Appendix]:%
\begin{equation}
\frac{a\cdot v}{a}\equiv \frac{a_{\mu }v^{\mu }}{a}=\frac{\gamma }{a}\left[
\left( a^{2}+\left\vert \mathbf{a}\right\vert ^{2}\right) ^{1/2}-\mathbf{a}%
\cdot \frac{d\mathbf{x}/dt}{c}\right] \geq 1,
\end{equation}%
and therefore%
\begin{equation}
\delta \left( \pm \frac{a\cdot v}{a}\right) =0,
\end{equation}%
and%
\begin{equation}
\delta ^{\prime }\left( \pm \frac{a\cdot v}{a}\right) =0.
\end{equation}%
Next, using Eqs. (20), (45), (50), (51) and (42), one obtains%
\begin{equation}
\square _{v}\phi _{2}^{\pm }(v)=\left( \frac{\rho _{0}}{\lambda _{0}}\right)
^{2}\phi _{2}^{\pm }(v).
\end{equation}%
Also, using Eqs. (31) and (36), Eq. (24) can be written as%
\begin{equation}
\square _{v}\phi _{2}(v)=\left( \frac{\rho _{0}}{\lambda _{0}}\right)
^{2}\phi _{2}(v).
\end{equation}%
Comparing Eq. (52) with Eq. (53), one concludes that $\phi _{2}^{\pm }(v)$
given by Eq. (42) solves Eq. (53). Next substituting Eqs. (41) and (42) in
Eq. (21), it follows that possible solutions to Eq. (17), representing the
positive and negative frequency modes of the quantum field are given by 
\begin{equation}
\phi ^{\pm }(x,v)=\phi _{0}e^{\pm ip\cdot x/\hbar }e^{\mp \frac{\rho _{0}}{%
\lambda _{0}}\frac{a\cdot v}{a}}\theta \left( \pm \frac{a\cdot v}{a}\right) ,
\end{equation}%
where%
\begin{equation}
\phi _{0}=\phi _{10}\phi _{20},
\end{equation}%
and, in accord with Eq. (33), the components of the particle four-momentum
are given by 
\begin{equation}
\left\{ p^{\mu }\right\} =\left\{ p^{0},\mathbf{p}\right\} =\left\{ \left(
m^{2}c^{2}+\left\vert \mathbf{p}\right\vert ^{2}\right) ,\mathbf{p}\right\} .
\end{equation}

The general solution $\phi (x,v)$ for a free relativistic scalar quantum
field is obtained by integrating over the spatial components of
four-momentum and four-acceleration of the invariant positive and negative
frequency modes $\phi ^{\pm }(x,v)$ of Eq.(54), including appropriate
particle creation and annihilation operators. It follows from Eqs. (54),
(49), and (27) that for the positive-frequency particles and the
negative-frequency antiparticles, respectively, nonvanishing support is
provided by positive and negative values of $\frac{a\cdot v}{a}$,
respectively. Thus the relativistic Lorentz-invariant scalar quantum field
for particles of mass $m$ is given by 
\begin{equation}
\begin{array}{c}
\phi (x,v)=2\int \frac{d^{3}\mathbf{p}}{N^{1/2}\left( 2\pi \hbar \right)
^{3/2}\left( 2p^{0}\right) ^{1/2}}\int \frac{d^{3}\mathbf{a}}{\left( 2\pi
\hbar \right) ^{3/2}\left( 2a^{0}\right) ^{1/2}}\left[ e^{-ip\cdot x/\hbar
}e^{-\frac{\rho _{0}}{\lambda _{0}}\frac{a\cdot v}{a}}\theta \left( \frac{%
a\cdot v}{a}\right) b(\mathbf{p,a})\right.  \\ 
+\;\left. e^{ip\cdot x/\hbar }e^{\frac{\rho _{0}}{\lambda _{0}}\frac{a\cdot v%
}{a}}\theta \left( -\frac{a\cdot v}{a}\right) b^{\dagger }(\mathbf{p,a})%
\right] .%
\end{array}%
\end{equation}%
Here, to recapitulate, $\theta (x)$ is the Heaviside step function, $x^{\mu }
$ is both the particle spacetime coordinate and the spacetime coordinate of
the measuring device that measures the quantum field, $p^{\mu }$ is the
particle four-momentum, $a^{\mu }$ is the particle four-acceleration, $%
v^{\mu }$ is the four-velocity of the measuring device, $\hbar $ is Planck's
constant divided by $2\pi $, $\rho _{0}=c^{2}/a_{M}$, $a_{M}$ is the maximal
proper acceleration given by Eq.\ (2), and $\lambda _{0}=\hbar /mc$. Also, $N
$ is a normalization constant, $b^{\dagger }(\mathbf{p,a})$ and $b(\mathbf{%
p,a})$\ are particle creation and annihilation operators for \ the spatial
component of four-momentum $\mathbf{p}$ and spatial component of
four-acceleration $\mathbf{a}$, and the following natural extensions of the
standard bosonic commutation relation are adopted: 
\begin{equation}
\left[ b(\mathbf{p,a}),b^{\dagger }(\mathbf{p}^{\prime }\mathbf{,a}^{\prime
})\right] =\delta ^{3}(\mathbf{p-p}^{\prime })\delta ^{3}(\mathbf{a-a}%
^{\prime }),
\end{equation}%
\begin{equation}
\left[ b(\mathbf{p,a}),b(\mathbf{p}^{\prime }\mathbf{,a}^{\prime })\right]
=0,
\end{equation}%
\begin{equation}
\left[ b^{\dagger }(\mathbf{p,a}),b^{\dagger }(\mathbf{p}^{\prime }\mathbf{,a%
}^{\prime })\right] =0.
\end{equation}%
Note that since the field is localized in spacetime, then even though the
proper acceleration $a$ is bounded by $a_{M}$, consistent with Eq. (37), the
magnitude of the spatial component of acceleration $\left\vert \mathbf{a}%
\right\vert \mathbf{\ }$of the particle may range from zero to infinity in
the integral. Also, one notes that\ according to Eqs. (2) and (11), $\rho
_{0}$ is of the order of the Planck length, and in the mathematical limit of
infinite limiting proper acceleration $a_{M}$, one has vanishing $\rho _{0}$%
, and Eq. (57) then reduces to the same form as a canonical
Lorentz-invariant scalar quantum field (as it must). In this sense, the
modification of the relativistic \ quantum field introduced here is
negligible, except at energies beyond the Planck energy $\left( \hbar
c^{5}/G\right) ^{1/2}$, as may be seen from the following.

Using Eqs. (2), (11), (36), and (49), one sees that both the positive and
negative frequency terms in Eq. (57) are proportional to 
\begin{equation}
\exp (-\frac{\rho _{0}}{\lambda _{0}}\left\vert \frac{a\cdot v}{a}%
\right\vert )=\exp \left[ -\frac{1}{2\pi \alpha }\frac{\gamma m}{m_{Pl}}%
\left\{ \left( 1+\left( \frac{|\mathbf{a|}}{a}\right) ^{2}\right) ^{1/2}-%
\frac{\mathbf{a}}{a}\cdot \frac{d\mathbf{x}/dt}{c}\right\} \right] ,
\end{equation}%
where $m$ is the rest mass of the particle, $m_{Pl}$ is the Planck mass,%
\begin{equation}
m_{Pl}=\left( \frac{\hbar c}{G}\right) ^{1/2},
\end{equation}%
and 
\begin{equation}
\gamma =\left( 1-\left\vert \frac{d\mathbf{x}/dt}{c}\right\vert ^{2}\right)
^{-1/2},
\end{equation}%
where $d\mathbf{x}/dt$ is the velocity of the measuring device relative to
the particle. For velocities of the measuring device much less than the
velocity of light, and for particles masses much less than the Planck mass,
Eq. (61) is for all practical purposes unity because $\rho _{0}$ is so small
(of the order of the Planck length), and Eq. (57) effectively reduces to the
canonical scalar quantum field.

\section{QUANTUM-TO-CLASSICAL TRANSITION}

Next, a possible implication of the limiting proper acceleration $a_{M}$ is
that for a particle with four-momentum $p^{\mu }$ and four-acceleration $%
a^{\mu }$, and for a measuring device with spacetime coordinates $x^{\mu }$
and four-velocity coordinates $v^{\mu }$ \ with respect to a particle in
Minkowski spacetime, the quantum state is given by%
\begin{equation}
\psi (x,v)=\left\langle 0|\phi (x,v)|p,\mathbf{a}\right\rangle .
\end{equation}%
Here $\left\vert 0\right\rangle $ is the vacuum state, $\left\vert p,\mathbf{%
a}\right\rangle $ is the state of the particle with four-momentum $p^{\mu }$
and four-acceleration $a^{\mu }$, and $\phi (x,v)$, given by Eq. (57), is
the scalar quantum field associated with the particle. From Eqs. (57), (61),
and (63), in the case in which the measuring device is at rest with respect
to the particle, namely $d\mathbf{x}/dt=0$ , it follows that the quantum
state is given by%
\begin{equation}
\psi (x,d\overrightarrow{x}/dt=0)=A\exp \left( -\frac{1}{2\pi \alpha }\frac{m%
}{m_{Pl}}\left( 1+\left( \frac{|\mathbf{a|}}{a}\right) ^{2}\right)
^{1/2}\right) \exp \left( -ip\cdot x/\hslash \right) ,
\end{equation}%
where $A$ is a normalization constant. Equation (65) can be expected to hold
for any bosonic or fermionic state, ignoring spin. For standard elementary
particle masses $\left( m/m_{Pl}\ll 1\right) $, the wave function Eq. (65)
reduces to the standard plane wave. But suppose that Eq. (65) is taken to
describe a macroscopic object containing many more than Avogadro's number of
atoms, in which case its mass $m$ is such that%
\begin{equation}
m\gg N_{A}m_{n},
\end{equation}%
where $N_{A}$ is Avogadro's Number, and $m_{n}$ is the mass of a nucleon.
Then according to Eqs. (65) and (66), in this simplified model, one obtains
for the quantum state $\psi _{mac}$ of this macroscopic many-body object:%
\begin{equation}
\psi _{mac}\ll A^{\prime }\exp \left( -\frac{1}{2\pi \alpha }N_{A}\frac{m_{n}%
}{m_{Pl}}\left( 1+\left( \frac{|\mathbf{a|}}{a}\right) ^{2}\right)
^{1/2}\right) \exp \left( -iN_{A}p\cdot x/\hslash \right) ,
\end{equation}%
in which $A^{\prime }$ is the normalization constant for $N_{A}$ particles.
Equivalently, one can write the state $\psi _{mac}$ as a product of $N_{A}$
copies of Eq. (65), since the exponents add. Substituting $N_{A}=$ $6\times
10^{23}$\ , $m_{n}=1.7\times 10^{-27}$kg , and $m_{Pl}=2.2\times 10^{-8}$ kg
in Eq. (67), the macroscopic many-body wave function $\psi _{mac}$ is seen
to be negligible. This suggests that, for all practical purposes, a
macroscopic object, such as a macroscopic measuring device, should not be
described by quantum mechanics, and instead is best described by classical
mechanics. This is consistent with Bohr's requirement that the macroscopic
measuring device be classical, and that it be clearly distinguished from the
microscopic quantum object being measured \cite{NB1}, \cite{VN1}, \cite{L1}.
It is also significant to note that in the present theory, the
quantum-to-classical transition is not sharp, but is gradual beyond
Avogadro's number of atoms. According to Eq. (67), the transition depends on
the total mass of the object being measured. Also noteworthy is that the
transition also depends on the universal gravitational constant $G$, which,
according to Eq. (62), appears in Eq. (67) . In this connection, one recalls
that Diosi introduced a theory of statistical mass localization with a
strength proportional to the gravitational constant $G$ \cite{D1}, \cite{L1}%
. It is also noteworthy that in other theories of the quantum-to-classical
transition, new universal constants were introduced, whereas in the present
theory there are no new independent universal constants. For example, in the
popular theory of G. C. Ghirardi, A. Rimini, and T. Weber (GRW theory), a
time scale was introduced for the rate of spontaneous localization in
spacetime, and also a length scale characterizing the narrowness of the
localization \cite{GRW1}, \cite{L1}. In the present theory, all of the
universal constants are standard, namely, $G$, $\hslash $, $c$, $N_{A}$, and
known particle masses. Also of note is that with current values of the
parameters in the GRW theory, significant tests of the theory could be
obtained with objects having more than $10^{8}$ nucleons, which is far less
than Avogadro's number \cite{L1}. It is also well to note that a bound on
the number of particles which can exist in a quantum entangled state, as in
a quantum computer, has earlier been conjectured on the basis of a
cosmological information bound in which more bits of information would be
needed to specify the state than can be accommodated in the observable
universe \cite{PD1}. In that conjecture, the required number of particles
for the transition to occur would be far less than Avogadro's number. It is
also well to note that the present theory in itself places no practical
upper bound on the size \ of a quantum computer register because Avogadro's
number is so large. Also, in the present theory, any quantum parallelism
involving the macroscopic every-day world, as in the Everett interpretation
of quantum mechanics, would be negligible \cite{E1}.

\section{CONCLUSION}

A new formulation of a relativistic scalar quantum field is here introduced
which generalizes the canonical theory to include the limiting proper
acceleration. This field is used to construct a simple model of a many-body
system, and it is argued that for a macroscopic object consisting of more
than Avogadro's number of atoms, any supposed quantum state of the object is
negligibly small, so that for all practical purposes, the object is best
described by classical mechanics. Thus, a new explanation is offered for the
quantum-to-classical transition and the absence of quantum superpositions of
macroscopic objects in the everyday world.

\section{APPENDIX}

\bigskip Equation (49) is to be proved. From Eq. (49, one obtains%
\begin{equation}
\frac{a\cdot v}{a}=\gamma \left[ \left( 1+\left( \frac{\left\vert \mathbf{a}%
\right\vert }{a}\right) ^{2}\right) ^{1/2}-\frac{\left\vert \mathbf{a}%
\right\vert }{a}\cdot \frac{d\mathbf{x}/dt}{c}\right] ,
\end{equation}%
or, equivalently,%
\begin{equation}
\frac{a\cdot v}{a}=\gamma F\left( \left\vert \mathbf{a}\right\vert /a\right)
,
\end{equation}%
where the function $F(x)$ is defined by%
\begin{equation}
F(x)\equiv \left( 1+x^{2}\right) ^{1/2}-\beta x\cos \theta ,
\end{equation}%
and 
\begin{equation}
\beta =\frac{|d\mathbf{x}/dt|}{c}<1,
\end{equation}%
\begin{equation}
\theta =\cos ^{-1}\left( \frac{\mathbf{a\cdot }d\mathbf{x}/dt}{\left\vert 
\mathbf{a}\right\vert |d\mathbf{x}/dt|}\right) ,
\end{equation}%
and using Eq. (71) in Eq. (63), one has 
\begin{equation}
\gamma =(1-\beta ^{2})^{-1/2}\geq 1.
\end{equation}%
Next, the function $F(x)$ has a minimum for 
\begin{equation}
0=\frac{\partial F}{\partial x}=x\left( 1+x^{2}\right) ^{-1/2}-\beta \cos
\theta ,
\end{equation}%
and it follows that the minimum is at%
\begin{equation}
x=x_{\min }=\frac{\beta \cos \theta }{\left( 1-\beta ^{2}\cos ^{2}\theta
\right) ^{1/2}}.
\end{equation}%
Then substituting Eq. (75) in Eq. (70), one obtains the minimum value of the
function $F(x)$, namely, 
\begin{equation}
F(x_{\min })=\left( 1-\beta ^{2}\cos ^{2}\theta \right) ^{1/2}.
\end{equation}%
Finally from Eqs. (69), (73), and (76), it follows that%
\begin{equation}
\frac{a\cdot v}{a}>\frac{\left( 1-\beta ^{2}\cos ^{2}\theta \right) ^{1/2}}{%
\left( 1-\beta ^{2}\right) ^{1/2}}>1,
\end{equation}

\end{document}